\documentclass[aps,pre,twocolumn,showpacs,longbibliography,superscriptaddress]{revtex4-1}

\usepackage{color}

\usepackage{ulem}

\usepackage{graphicx}

\begin{document}
\newcommand{\ud}{{\mathrm d}}
\newcommand{\sech}{\mathrm{sech}}

\title{Brillouin propagation modes of cold atoms in dissipative optical lattices}
\author{David Cubero}
\email[]{dcubero@us.es}
\affiliation{Departamento de F\'{\i}sica Aplicada I, Escuela Polit\'ecnica Superior, Universidad de Sevilla, Calle Virgen de \'Africa 7, 41011 Sevilla, Spain}


\begin{abstract}
An exact series expansion of the average velocity of cold atoms in dissipative optical lattices under probe driving,  based on the amplitudes of the excited atomic density waves, is derived from the semiclassical equations for the phase space densities of the Zeeman ground-state sublevels. This expansion permits the identification of the precise contribution to the current of a propagating atomic wave for the specific driving, as well as providing the threshold for the transition into the regime of infinite density.
\end{abstract}

\maketitle

\section{Introduction}

In the last 40 years, laser cooling of atoms to very low temperatures \cite{schreck21} has been a crucial experimental achievement in quantum and solid state physics,  with several Nobel prices awarding the development of techniques in this subject. Among them, cold atoms in dissipative optical lattices \cite{grynreview} still play an important role, and Sisyphus cooling in particular is still being widely used to cool atoms below the Doppler temperature limit.

Cold atoms offer an invaluable setup to study directed transport, and many experimental realizations of dissipative \cite{cecile99,schiavoni2003,gommers2005,gommers2006,jones2007,wichol12} and nondissipative \cite{salger2009,denisov15} ratchets  \cite{hanmar09,cubren16} have been demonstrated on them. Additionally, they are also known to exhibit unsual transport behaviour beyond Boltzmann–Gibbs statistical mechanics \cite{lutren13}, usually referred as a regime of infinite density  \cite{kessler10,hol15,barkai21,barkai22} in which the probability distributions are non-normalizable due to the broken ergodicity.

This paper focused in cold atoms in standard setups of optical lattices associated with Sisyphus cooling.
They are created by counterpropagating laser beams, and can be tuned experimentally with a high level of control. Here they are studied theoretically at the level of the semiclassical equations for the atomic phase space  densities, which allows a detailed description of the Sisyphus mechanism. 

Special attention is paid to atomic transport. A perturbing probe beam is introduced \cite{gryn96,grynreview,renzoni02a} to break the system symmetries \cite{fabio07,cubren16} and put the atoms into directed motion.
The probe perturbation excites atomic density waves, which are referred to as Brillouin propagating modes \cite{renzoni02a} due  to the resemblance to acoustic waves rippling through a dense fluid, even though in the present case the optical lattice is dilutely occupied and the mode is sustained in a medium of non-interacting particles due to the interaction with the laser beams. 

Previous research has focused mostly on one propagating mode, the one with same frequency and wave number as the propagating perturbation, ignoring the effect of other excited modes, propagating or not. Here we address this issue by deriving a series expansion of the current on the amplitudes of the excited atomic density waves. Additionally, the expansion directly reveals a singularity that is identified with the threshold to the regime of infinity density. The threshold values are in agreement with previous analytical results \cite{kessler10,hol15} obtained from a simplified, approximate Fokker-Planck equation \cite{cohentan92} based on space and Zeeman sublevel averaging of the semiclassical equations---thus with uncontrolled approximations.   The method presented here does not make such approximations, being based on the semiclassical equations it takes explicitly into account the microscopic
origin of Sisyphus cooling, 
thus providing desirable support for previous results on infinity densities.

The paper is organized as follows. In Sec.~\ref{sec:models} the system models studied are described in detail. The new method, relating the current and higher moments to the atomic density modes, is presented in Sec.~\ref{sec:analytic}. The application of the method to the system models, its analytical results and numerical validation, is in Sec.~\ref{sec:newtheor}. The transition to the regime of infinity density, coming out naturally from a singularity in the analytical results, is discussed in in Sec.~\ref{sec:inf}.  Finally, Sec.~\ref{sec:con} ends with the main conclusions.

\section{System models}
\label{sec:models}

First, we consider the simplest model of a dissipative optical lattice with Sisyphus cooling, a one-dimensional (1D) system generated by atoms with a transition $J_g= 1/2 \rightarrow J_e=3/2$, mass $m_a$, illuminated by two counter-propagating laser fields with orthogonal linear polarizations. This setup generates a 1D optical lattice,  so-called lin$\perp$lin configuration \cite{grynberg2001}. The atom, in the Zeeman sub-level of the atomic ground-state $+$ or $-$,  experiences the optical potential
\begin{equation}
U_{\pm}(x)= \frac{U_0}{2}[-2\pm \cos(2k_l x)],
\label{eq:upm1D}
\end{equation}
 where $x$ is the laser beam propagation axis, $k_l$ the laser field wave vector, and $U_0$ the optical lattice depth. 

Further details of the dissipative optical lattice are given, in the semi-classical approximation for weak laser intentities, by the following  \cite{petsas99}  coupled Fokker-Planck equations for the phase space density $P_{\pm}(x,p,t)$ of atoms in the ground state Zeeman sublevel $|\pm\rangle=|J_g=1/2,M_g=\pm1/2\rangle$ at the position $x$ with momentum $p$,
\begin{eqnarray}
\left[ \frac{\partial}{\partial t} + \frac{p}{m_a}\frac{\partial}{\partial x} 
-U_{\pm}^\prime(x)\frac{\partial}{\partial p}+F_{\pm}(x,t)\frac{\partial}{\partial p}\right] P_{\pm}= \nonumber\\
-\gamma_{\pm}(x) P_{\pm}+\gamma_{\mp}(x) P_{\mp} 
+\frac{\partial^2}{\partial p^2}\left[D_0 P_{\pm}\right],
\label{eq:fp}
\end{eqnarray}
where $\gamma_{\pm}$ is the transition rate between the ground state sublevels,  $D_0$ is a noise strength describing the random momentum jumps that result from the interaction with the photons, 
and $F_\pm(x,t)$ is generally a non-conservative force---coming, for example, from radiation pressure, or, as a second example,  being an arbitrary time-dependent driving force $F(t)$ that can be generated by phase modulating one the lattice beams \cite{ren09}.

Equation (\ref{eq:fp}) is complemented by the normalization condition, 
\begin{equation}
\int dx \int \!dp\left[ P_{-}(x,p,t) + P_{+}(x,p,t) \right] =1.
\label{eq:norm:fp}
\end{equation}

A quantity of special interest is the average atomic current, which measures the directed motion, being defined as
\begin{eqnarray}
&\langle v \rangle = \lim_{t\rightarrow\infty} \frac{\langle [x(t)-x(0)] \rangle }{t} = \nonumber\\
 &\lim_{t\rightarrow\infty}  \frac{ \int_0^t \!\! dt' \int \!\! dx \int \!\! dp \, \frac{p}{m_a}[P_+(x,p,t')+P_-(x,p,t')]}{t}.
\label{eq:currdef}
\end{eqnarray}

\subsection{1D lin$\perp$lin setup}
In the 1D lin$\perp$lin configuration without driving there is no radiation pressure, i.e. $F_\pm(x,t)=0$, the optical potential is given by (\ref{eq:upm1D}), with 
\begin{equation}
U_0=-2\hbar \Delta'/3>0,
\label{eq:u0:def}
\end{equation}
where $\Delta'$ is the light-shift per beam for a closed transition having a Clebsch-Gordan coefficient equal to unity. Furthermore, the transition rates between the internal states are given by
\begin{equation}
\gamma_{\pm}(x)=g_0\pm g_1\cos(k_0 x),
\label{eq:rates}
\end{equation}
where $k_0=2k_l$,  $g_0=\Gamma^\prime/9$ and $g_1=g_0$, with $\Gamma^\prime$ being the photon scattering rate per lattice beam.

Note that in writing (\ref{eq:fp}) we are not explicitly considering an extra diffusive term that comes up in the semiclassical approximation \cite{petsas99}, $\partial^2 D_{\mp\pm} P_{\mp}/\partial p^2$, which describes a further momentum kick when the transitions take place,  and has to be corrected ad hoc to avoid artificial singuralities produced by the semi-classical approximation \cite{hol15}. When corrected, the effect of that term is nevertheless small, and commonlly neglected, unless shallow optical potentials are considered \cite{hol15}---more about this issue in Sec.~\ref{sec:inf}.  

In addition, we are not explicitly considering any state or spatial dependence of the noise coefficient $D_0$, because their effect is observed to be small in the simulation results reported in this paper.
Neglecting $D_{\mp\pm}$ and taking the space average of the remaining diffusion coefficient in the original semi-classical equations \cite{petsas99} yields  $D_0=35\hbar^2k_l^2\Gamma^\prime/90$.

The application of an additional weak probe beam generically produces extra small contributions to all the functions above: the optical potential, radiation pressure forces, transition rates and noise terms, though the most relevant contribution usually is the one in the optical potential, being the others negligible. We consider here a probe beam that is polarized parallel to the 1D counter-propagating lattice beam, which yields the following optical potential
\begin{eqnarray}
U_{\pm}(x)= \frac{U_0}{2}[-2\pm \cos(k_0 x)  + \varepsilon_p \cos(k_0 x-\delta_p t+\phi_p)  ], \nonumber\\
\label{eq:upm1D:v1b}
\end{eqnarray}
where $\varepsilon_p=2E_p/E_0$ is twice the ratio between the electric field of the probe $E_p$ to that of the underlying optical lattice $E_0$, $\delta_p$ is the probe frequency detuning and $\phi_p$ the probe phase.

\subsection{3D lin$\perp$lin setup}
 In addition to the above 1D lin$\perp$lin configuration, we will study also the one-dimensional system model that arises in the standard  3D-lin$\perp$lin configuration \cite{grynreview},
 after neglecting movement in the two directions perpendicular to the direction of interest, usually taken as the $x$-axis \cite{gryn96,renzoni02b,renzoni03,renzoni02a,grynpra2003}. A weak $y$-polarized probe is added in the $z-$direction \cite{renzoni02a,grynpra2003}, yielding an extra term in the optical potential which is the superposition of two sinusoidal potentials travelling in the $x$-direction, each with opposite velocity and similar shape as in the previous 1D system model. More specifically, by formally considering $y=z=0$ and one of the travelling probe drives, the optical potential in each sublevel of the ground state is given by  \cite{grynreview,grynpra2003}
\begin{eqnarray}
U_\pm(x)=\frac{U_0}{2}\Big[-\frac{3}{2}-\frac{1}{2}\cos(2 k_0 x) \pm  \cos(k_0 x)   \nonumber \\
+ \varepsilon_p\cos(k_0 x -\delta_p t+\phi_p)
\Big],
\label{eq:Upm:3D}
\end{eqnarray}
with transition probability rates between them given by
\begin{equation}
\gamma_{\pm}(x)=g_0\pm g_1\cos(k_0 x) + g_2\cos(2 k_0 x) ,
\label{eq:gammapm}
%
\end{equation}
where $k_0=k_l\sin\theta_x$,  
$U_0=-16\hbar \Delta'_0/3 $,  
 $\Delta'_0$($<0$) is the light-shift per lattice field,
 $g_0=2\Gamma_s/3$, $g_1=8\Gamma_s/9$, $g_2=2\Gamma_s/9$,
and $\Gamma_s$ is the photon scattering rate per lattice beam, and now $ \varepsilon_p=E_p/(2E_0)$.
 Like before, for the sake of simplicity, we are neglecting the probe contribution to the transition rates, the radiation forces, and noise terms, being observed their effect to be small in the simulations. The equation to solve is also  (\ref{eq:fp}),  with  the noise strength $D_0=5\hbar^2k_0^2\Gamma_s/18$.

\section{Fourier mode theory}
\label{sec:analytic}

We develop in this section a theory that will allow us to visualize the atomic density modes excited by the probe and their precise contribution to the current. 

First, the atomic Fourier modes are defined from the phase space density $P_{\pm}(z,p,t)$  by means of the following Fourier transform
\begin{equation}
\mathcal{P}^\pm_{\omega,k,q}=\frac{1}{T_p}\int_0^{T_p}\!\!\! dt \, e^{-i\omega t}\int \!dx \,  e^{i k x}\int \!dp \,e^{i pq/\hbar}  P_{\pm}(x,p,t),
\label{eq:pfourier:def}
\end{equation}
where $T_p=2\pi/\delta_p$ is the time period introduced by the probe.  The fact that the driving probe is time periodic allows us to focus only on solutions which has the same periodicity, i.e. to 
\begin{equation}
\omega=l \delta_p,
\label{eq:om:discrete}
\end{equation}
 with $l$ integer---thus neglecting transitory dependencies on a specific initial condition, which are expected to die out after a transient time interval, leaving the periodic solution.  
In addition, real phase space densities implies
\begin{equation}
\mathcal{P}^\pm_{-\omega,-k,-q}=\left( \mathcal{P}^{\pm}_{\omega,k,q}\right)^*,
\label{eq:pfourier:def:real}
\end{equation}
where $*$ denotes complex conjugate.

Now, let us consider a generic 1D setup determined by (\ref{eq:fp}) and (\ref{eq:norm:fp}), with a periodic optical potential
\begin{equation}
U_\pm(x+2\pi/k_0)=U_\pm(x),
\end{equation}
 and space periodic transition rates, 
\begin{equation}
\gamma_\pm(x+2\pi/k_0)=\gamma_\pm(x), 
\end{equation}
and time and space periodic driving,
\begin{equation}
F_\pm(x+2\pi/k_0,t)=F_\pm(x,t+2\pi/\delta_p)=F_\pm(x,t),
\label{eq:force:period}
\end{equation}
for all $x, t$.

Using (\ref{eq:pfourier:def}), the Fokker-Planck Eq.~(\ref{eq:fp}) is transformed into
\begin{eqnarray}
& i\omega \mathcal{P}^\pm_{\omega,k,q}
 -\frac{\hbar k}{m_a}\frac{\partial}{\partial q}\mathcal{P}^\pm_{\omega,k,q}  \nonumber\\
 & -\frac{iq}{\hbar}\Big(    
          \sum_n\mathcal{F}^{(0)\pm}_n \mathcal{P}^\pm_{\omega,k-nk_0,q}  
        + \sum_{l,m} \mathcal{F}^{p\pm}_{l,m}   \mathcal{P}^\pm_{\omega-l\delta_p,k-mk_0,q}
           \Big) = \nonumber \\
&           -\sum_n \Big(   \gamma_n^{\pm} \mathcal{P}^\pm_{\omega,k-nk_0,q}   - 
                                \gamma_n^{\mp} \mathcal{P}^\mp_{\omega,k-nk_0,q} \Big) \nonumber\\
 &                               -\frac{q^2}{\hbar^2} D_0  \mathcal{P}^\pm_{\omega,k,q} \,\,,
\label{eq:fp:new}
\end{eqnarray}
where we have taken advantage of the mentioned periodicity, allowing us to write
\begin{eqnarray}
-\frac{\partial U_\pm}{\partial x} = \sum_n \mathcal{F}^{(0)\pm}_n e^{-i n k_0 z}, \label{eq:F0def}\\
F_\pm(z,t) = \sum_{l,m} \mathcal{F}^{p\pm}_{l,m} e^{-i ( m k_0 z -l\delta_p t)`},
\label{eq:driving:pot}
\end{eqnarray}
and similarly to the transition rates
\begin{equation}
\gamma_\pm(z) = \sum_n \gamma^{\pm}_n e^{-i n k_0 z}.
\label{eq:gamdef}
\end{equation}
In the problems considered here, there are no force bias, i.e.
\begin{equation}
\mathcal{F}^{(0)\pm}_0=\mathcal{F}^{p\pm}_{0,0}=0,
\label{eq:nobias}
\end{equation}
and the states are symmetric:
\begin{equation}
\gamma_0^\pm=\gamma_0.
\label{eq:g0}
\end{equation}

From (\ref{eq:pfourier:def}), it is clear the normalization condition  (\ref{eq:norm:fp}) is translated into
\begin{equation}
\mathcal{P}^-_{0,0,0}+\mathcal{P}^+_{0,0,0}=1.
\label{eq:norm:fp:new}
\end{equation}

Once $\mathcal{P}^\pm_{\omega,k,q}$ is known, the phase space density is obtained back by means of the inverse transform,
 \begin{equation}
P_{\pm}(x,p,t)=\frac{1}{4\pi^2\hbar}\sum_l \int \!dk \int \!dq \, e^{i (-k x+tl\delta_p )}e^{-i pq/\hbar}  \mathcal{P}^\pm_{l\delta_p,k,q}.
\end{equation}
We are specially interested in the marginal probability $P_{\pm}(x,t)=\int dp P_\pm(x,p,t)$, which can be obtained from $ \mathcal{P}^\pm_{\omega,k,q}$ with $q=0$, 
 \begin{equation}
P_{\pm}(x,t)= \frac{1}{2\pi}\sum_l \int \!dk \,  \mathcal{P}^\pm_{l\delta_p,k,0} e^{-i (k x- t l \delta_p)}.
\label{eq:prob:mar}
\end{equation}
The normalization condition (\ref{eq:norm:fp:new}) and the equations of motion (\ref{eq:fp:new})  only requires excitation of a discrete set of possible wave number values,
\begin{equation}
k=n k_0
\label{eq:k:discrete}
\end{equation}
where $n$ is an integer.

Note equation (\ref{eq:prob:mar}) expresses the atomic density as a sum of plane waves, atomic density modes, each moving with velocity $(l/n)\delta_p/k_0$. The case $(l,n)=(1,1)$ has been explicitly referred to as a Brillouin-like propagation mode \cite{gryn96,grynreview,renzoni02a}, because of the analogy with the resonances produced by light scattering on propagative modes in fluids, such as sound waves. 

Assuming, like in (\ref{eq:pfourier:def}), that the time $t=0$ is after all the transients have already died out, the atomic current (\ref{eq:currdef}) can be written in terms of the mode amplitudes as
\begin{eqnarray}
\langle v \rangle &=& \frac{\delta_p}{2\pi} \int_{0}^{2\pi/\delta_p} \!\! dt \int \!\! dx \int \!\! dp \, \frac{p}{m_a}[P_+(x,p,t)+P_-(x,p,t)] \nonumber\\
&=& \lim_{q\rightarrow 0}\frac{\hbar \delta_p }{i m_a 2\pi}\frac{\partial}{\partial q} \int_0^{2\pi/\delta_p} \!\! dt \int dx\int dp \, e^{i(k x-\omega t+pq/\hbar)} \nonumber\\
&{ }&\quad \quad\times [P_+(x,p,t)+P_-(z,p,t)] \Big|_{\omega=k=0} \nonumber\\
&=& \frac{\hbar}{i m_a}\lim_{q\rightarrow 0}\left( \frac{\partial \mathcal{P}^+_{0,0,q}}{\partial q}+\frac{\partial \mathcal{P}^-_{0,0,q}}{\partial q} \right).
\label{eq:currdef:four}
\end{eqnarray}

The equations (\ref{eq:fp:new}) and (\ref{eq:norm:fp:new}) can be used to find $\partial \mathcal{P}^\pm_{0,0,0}/\partial q$ in terms of the mode amplitudes $\mathcal{P}^\pm_{\omega,k,0}$, thus providing the exact contribution of each excited mode to the atomic current,  yielding one of the central analytical results of this paper, in which the current  (\ref{eq:currdef}) is written as
\begin{equation}
\langle v \rangle  =\sum_{l,n} v[l,n] 
=\sum_{l,n} v^+_{l,n}\mathcal{P}^+_{l\delta_p,nk_0,0} + v^-_{l,n}\mathcal{P}^-_{l\delta_p,nk_0,0},
\label{eq:vmodes}
\end{equation}
with explicit analytical expressions for the coefficients $ v^\pm_{l,n}$ for the specific system at hand, as reported in Sec.~\ref{sec:newtheor}.

The same procedure can be applied to compute any moment of the velocity distribution, $\langle v^n \rangle $, with $n$ a positive integer, since following similar steps as in (\ref{eq:currdef:four}) one easily obtains
\begin{eqnarray}
\langle v^n \rangle = \left(\frac{\hbar}{i m_a}\right)^n \lim_{q\rightarrow 0}\left( \frac{\partial^n \mathcal{P}^+_{0,0,q}}{\partial q^n}+\frac{\partial^n \mathcal{P}^-_{0,0,q}}{\partial q^n} \right),
\label{eq:moment}
\end{eqnarray}
with $q$-derivatives that can be found in terms of the amplitudes of the atomic waves in a similar fashion as before.

\subsection{Basic symmetry}
Due to the vectorial nature of both the driving force $F_\pm(x,t)$ and the current $\langle v\rangle$, the inversion transformation $x\rightarrow-x$ yields an inverted current \cite{cubren_prl2016}, a fact which is behind the ---easy to verify--- statement that the solution of the problem with conservative force
\begin{equation}
\widehat{\mathcal{F}}^{(0)\pm}_n = -\mathcal{F}^{(0)\pm}_{-n},
\end{equation}
and driving force
\begin{equation}
\widehat{\mathcal{F}}^{p\pm}_{l,m}= -\mathcal{F}^{p\pm}_{l,-m},
\label{eq:oppforce}
\end{equation}
is just
\begin{equation}
\widehat{\mathcal{P}}^\pm_{\omega,k,q}=\mathcal{P}^\pm_{\omega,-k,-q},
\end{equation}
which, by virtue of (\ref{eq:currdef:four}), yields
\begin{equation}
\langle \widehat{v}\rangle= -\langle v\rangle.
\label{eq:vinver}
\end{equation}
Note if the potential is spatially symmetric, then 
\begin{equation}
\mathcal{F}^{(0)\pm}_n = -\mathcal{F}^{(0)\pm}_{-n},
\label{eq:spasymm}
\end{equation}
and to produce a current we need a symmetry breaking probe.

\subsection{Sketch of the calculation of the current in terms of mode amplitudes}
\label{sec:sketch}
From (\ref{eq:currdef:four}), the calculation of the current is reduced to find an expression for $\partial \mathcal{P}^\pm_{0,0,0}/\partial q$.  However, this is not a trivial task, because the term proportional to $\partial/\partial q$ in (\ref{eq:fp:new}) vanishes for $k=0$. 

To proceed, we focus in  (\ref{eq:fp:new})  for $\omega=k=0$, and Taylor expand the dependency on $q$, having for small $q$,
\begin{equation}
\mathcal{P}^\pm_{0,0,q} = \mathcal{P}^\pm_{0,0,0} + q \frac{\partial \mathcal{P}^\pm_{0,0,0}}{\partial q}+  \frac{q^2}{2} \frac{\partial^2 \mathcal{P}^\pm_{0,0,0}}{\partial q^2}+\ldots
\label{eq:taylor}
\end{equation}
After summing the $-$ and $+$ forms of Eq.~(\ref{eq:fp:new}) for $\omega=k=0$ and inserting the expansion (\ref{eq:taylor}), we find, in each order, with $n$ a non-negative integer,
\begin{eqnarray}
&0& =     \sum_n \left( \mathcal{F}^{(0)+}_{n'} \mathcal{P}^+_{0,-n'k_0,0}  + 
                        \mathcal{F}^{(0)-}_{n'} \mathcal{P}^-_{0,-n'k_0,0} \right)  \nonumber\\
 & &    + \sum_{l,m} \left( \mathcal{F}^{p+}_{l,m}   \mathcal{P}^+_{-l\delta_p,-mk_0,0} 
                             + \mathcal{F}^{p-}_{l,m}   \mathcal{P}^-_{-l\delta_p,-mk_0,q} \right),  \\
& &   \frac{(n+1)D_0}{i\hbar}    \left( \frac{\partial^n \mathcal{P}^+_{0,0,0}}{\partial q^n}+\frac{\partial^n \mathcal{P}^-_{0,0,0}}{\partial q^n} \right)   =   \nonumber \\
 & &    \sum_{n'} \left( \mathcal{F}^{(0)+}_{n'} \frac{\partial^{n+1} \mathcal{P}^+_{0,-n'k_0,0}}{\partial q^{n+1}}  + 
                              \mathcal{F}^{(0)-}_{n'} \frac{\partial^{n+1} \mathcal{P}^-_{0,-n'k_0,0}}{\partial q^{n+1}} \right)  \nonumber\\
 & &      + \sum_{l,m} \Big( \mathcal{F}^{p+}_{l,m}   \frac{\partial^{n+1} \mathcal{P}^+_{-l\delta_p,-mk_0,0}}{\partial q^{n+1}}  \nonumber\\
 & &           \quad  \quad  \quad             + \mathcal{F}^{p-}_{l,m}   \frac{\partial^{n+1} \mathcal{P}^-_{-l\delta_p,-mk_0,0}}{\partial q^{n+1}} \Big).               \label{eq:curr:sol}       \nonumber\\                   
\end{eqnarray}
Eq.~(\ref{eq:curr:sol}) with $n=1$ facilitates the task to calculate the current, because, by means of (\ref{eq:currdef:four}), it provides a useful expression in terms of derivatives with respect to $q$ of amplitudes with either $k\ne0$ or $\omega\ne0$. These derivatives can be calculated by using the following equations, obtained by using the Taylor expansion 
\begin{equation}
\frac{ \partial^{n'} \mathcal{P}^\pm_{\omega,k,q}}{\partial q^{n'}} =\frac{\partial^{n'} \mathcal{P}^\pm_{\omega,k,0} }{\partial q^{n'}} + q \frac{\partial^{1+n'} \mathcal{P}^\pm_{\omega,k,0}}{\partial q^{1+n'}}+  \frac{q^2}{2} \frac{\partial^{2+n'} \mathcal{P}^\pm_{\omega,k,0}}{\partial q^{2+n'}}+\ldots 
\label{eq:taylor2}
\end{equation}
with $n'$ a non-negative integer, in (\ref{eq:fp:new}), 
\begin{eqnarray}
&\frac{\hbar k}{m_a}\frac{\partial  \mathcal{P}^\pm_{\omega,k,0}  }{\partial q} &= i\omega  \mathcal{P}^\pm_{\omega,k,0}    \nonumber \\
 & &+ \sum_n \left( \gamma^\pm_n \mathcal{P}^\pm_{\omega,k-nk_0,0}  -
                       \gamma^\mp_n \mathcal{P}^\mp_{\omega,k-nk_0,0} \right),  
                       \label{eq:fp_sol:1}\\
 &\frac{\hbar k}{m_a}\frac{\partial^2  \mathcal{P}^\pm_{\omega,k,0}  }{\partial q^2} &= i\omega  \frac{\partial \mathcal{P}^\pm_{\omega,k,0}}{\partial q}  
-\frac{i}{\hbar} \sum_n \mathcal{F}^{(0)\pm}_n \mathcal{P}^\pm_{\omega,k-nk_0,0}   \nonumber \\
 &  & -\frac{i}{\hbar}     \sum_{l,m}        \mathcal{F}^{p\pm}_{l,m} \mathcal{P}^\pm_{\omega-l\delta_p,k-mk_0,0}   \nonumber\\
    & &+ \sum_n \left( \gamma^\pm_n \frac{\partial \mathcal{P}^{\pm}_{\omega,k-nk_0,0} }{\partial q} -
                       \gamma^\mp_n \frac{\partial \mathcal{P}^{\mp}_{\omega,k-nk_0,0} }{\partial q}\right). \nonumber\\
            \label{eq:fp:sol:2}                 
\end{eqnarray}
The direct use of (\ref{eq:fp_sol:1})--(\ref{eq:fp:sol:2}) allows us to calculate most terms, except $\partial \mathcal{P}^{\mp}_{\omega,0,0}/\partial q$, with $\omega\ne0$. In that case, that first derivative can also be found from (\ref{eq:fp:sol:2}), but it requires linear inversion of a matrix of range $2\times2$, because similar first derivative terms also appears in the right hand side.  Performing that calculation we obtain
\begin{eqnarray}
&\frac{\partial  \mathcal{P}^\pm_{\omega,0,0}  }{\partial q} &= \frac{1}{2\gamma_0 i \omega-\omega^2}\Bigg[
  \frac{\gamma_0 i}{\hbar} \Big[   \nonumber\\
& &   \sum_n \left( \mathcal{F}^{(0)+}_n \mathcal{P}^+_{\omega,-nk_0,0}  + 
                        \mathcal{F}^{(0)-}_n \mathcal{P}^-_{\omega,-nk_0,0} \right)    \nonumber\\
     &   &+ \sum_{l,m} \left( \mathcal{F}^{p+}_{l,m}   \mathcal{P}^+_{\omega-l\delta_p,-mk_0,0} 
                             + \mathcal{F}^{p-}_{l,m}   \mathcal{P}^-_{\omega-l\delta_p,-m k_0,0} \right)                       
  \Big]   \nonumber \\
& &   -\frac{\omega}{\hbar} \Big[
 \sum_n \mathcal{F}^{(0)\pm}_n \mathcal{P}^\pm_{\omega,k-nk_0,0} 
 +\sum_{l,m} \mathcal{F}^{p\pm}_{l,m}\mathcal{P}^\pm_{\omega-l\delta_p,k-mk_0,0} 
     \Big]   \nonumber \\  
 & &- i\omega \sum_{n\ne0} \left( \gamma^\pm_n \frac{\partial \mathcal{P}^\pm_{\omega,-nk_0,0}}{\partial q}  -
                       \gamma^\mp_n \frac{\partial \mathcal{P}^\mp_{\omega,-nk_0,0}}{\partial q} \right) \Bigg].
                       \label{eq:fp_sol:dp}
\end{eqnarray}

Applying the above steps, it is then straightforward to arrive to an explicit expression for the current $\langle v\rangle$ ---or an arbitrary moment $\langle v^n\rangle$ using similar steps--- in the form (\ref{eq:vmodes}) for a particular system. The application to a system with Sisyphus cooling, such as those presented in Sec.~\ref{sec:models} reveals the presence of the regime of infinity density, as discussed in Sec.~\ref{sec:inf}.

\subsection{Calculation of the mode amplitudes in the computer simulation}
Numerical solutions of the equation (\ref{eq:fp}) are obtained by generating a large number of individual atomic trajectories \cite{petsas99} using a stochastic algorithm \cite{kloeden}.

For convenience,  let us denote the atomic mode amplitudes for $q=0$ as
\begin{equation}
P_\pm[l,n]=\mathcal{P}^\pm_{l\delta_p,n k_0,0}.
\end{equation}
They are the Fourier transform, Eq.~(\ref{eq:pfourier:def}), of the atomic density $P_\pm(x,t)$,
\begin{eqnarray}
P_\pm[l,n]=\frac{\delta_p}{2\pi}\int_0^{2\pi/\delta_p}\!\!\! dt \, e^{-i l \delta_p t}\int \!dx \,  e^{i n k_0 x}   P_{\pm}(x,t). \nonumber \\
= \lim_{l'\rightarrow\infty}\frac{\delta_p}{2\pi l'}\int_0^{2\pi l'/\delta_p}\!\!\! dt \, e^{-i l \delta_p t}\int \!dx \,  e^{i n k_0 x}   P_{\pm}(x,t).  \nonumber \\
\label{eq:pfourier:def2}
\end{eqnarray}
The atomic density can be expressed in terms of atomic trajectories $x^{\pm}_j(t)$ using the Dirac delta function
\begin{equation}
P_{\pm}(x,t)= \frac{1}{N}\sum_{j=1}^N  \delta[x-x^{\pm}_j(t)] .
\label{eq:dens:dirac}
\end{equation}
Thus, inserting (\ref{eq:dens:dirac}) into (\ref{eq:pfourier:def2}) yields
\begin{equation}
P_\pm[l,n]= \lim_{l'\rightarrow\infty} \frac{\delta_p}{2\pi l' N}\sum_{j=1}^{N}\int_0^{2\pi l'/\delta_p}\!\!\! dt \,  e^{i [n k_0 x^{\pm}_j(t) - l \delta_p t] }.
\label{eq:pfourier:def3}
\end{equation}
Since in the simulation individual atomic trajectories are generated, Eq.~(\ref{eq:pfourier:def3}) can be readily implemented to compute the atomic mode coefficients, involving an average over many atomic trajectories and many probe periods. At a particular time, the atom is in a definite state $+$ or $-$, and thus, it only contributes to the corresponding density via (\ref{eq:pfourier:def3}).

\section{Application to the system models}
\label{sec:newtheor}

Our goal is to express the current as (\ref{eq:vmodes}), i.e. an expansion of the atomic mode amplitudes 
$P_\pm[l,n]$.

In the system 1D lin$\perp$lin, the force and rate constants (\ref{eq:F0def})  and (\ref{eq:gamdef}), defining the optical potential (\ref{eq:upm1D}) and transition rates (\ref{eq:rates}), respectively, are given by
\begin{eqnarray}
\mathcal{F}^{(0)+}_{\pm1}=\mp\frac{F_0}{2i}, 
\quad  \mathcal{F}^{(0)-}_{\pm1}=-\mathcal{F}^{(0)+}_{\pm 1}  \label{eq:F0:1}\\
\gamma_{\pm 0}^\pm=g_0, \quad \gamma_{\pm1}^+= \frac{g_1}{2}, \quad  \gamma_{\pm1}^-=-\gamma_{\pm1}^+. \label{eq:g0:1}
\end{eqnarray}
with $F_0=k_0 U_0/2$. 

In the system 3D-lin$\perp$lin, Eqs.~(\ref{eq:Upm:3D})--(\ref{eq:gammapm}), in addition to  (\ref{eq:F0:1})--(\ref{eq:g0:1}),  there are the following coefficients
\begin{eqnarray}
\mathcal{F}^{(0)+}_{\pm2}=\pm\frac{F_0}{2i}, 
\quad  \mathcal{F}^{(0)-}_{\pm2}=\mathcal{F}^{(0)+}_{\pm 2}  \label{eq:F0:2}\\
 \gamma_{\pm2}^+= \frac{g_2}{2}, \quad  \gamma_{\pm2}^-=\gamma_{\pm2}^+. \label{eq:g0:2}
\end{eqnarray}

In both cases, the probe addition to the potential has the same form, and the driving force coefficients (\ref{eq:driving:pot}) are given by
\begin{equation}
\mathcal{F}^{p\pm}_{1, 1} = -\frac{F_p e^{-i\phi_p}}{2i}, \quad 
\mathcal{F}^{p\pm}_{-1, -1} = \frac{F_p e^{+i\phi_p}}{2i},
\label{eq:fp:v1b}
\end{equation}
where $F_p=U_0\varepsilon_p k_0/2$.

It can be checked that the solution, of both systems, possesses the following atomic state symmetry
\begin{equation}
\mathcal{P}^-_{l\delta_p,n k_0,q}= (-1)^{n-l} \mathcal{P}^+_{l\delta_p,n k_0,q}.
\label{eq:pminus:b}
\end{equation}
The proof involves verifying that (\ref{eq:pminus:b}) provides indeed a valid solution of  (\ref{eq:fp:new})---because of the specific way the sign of each term is changed in each atomic state---and invoking the uniqueness of the solution.  This general, simple relation for  the mode amplitudes $\mathcal{P}^-$ in terms of the coefficients in the other state, $\mathcal{P}^+$ allows us to focus only on the densities of one atomic sub-level.

\begin{figure}[t]
\includegraphics[width=8cm]{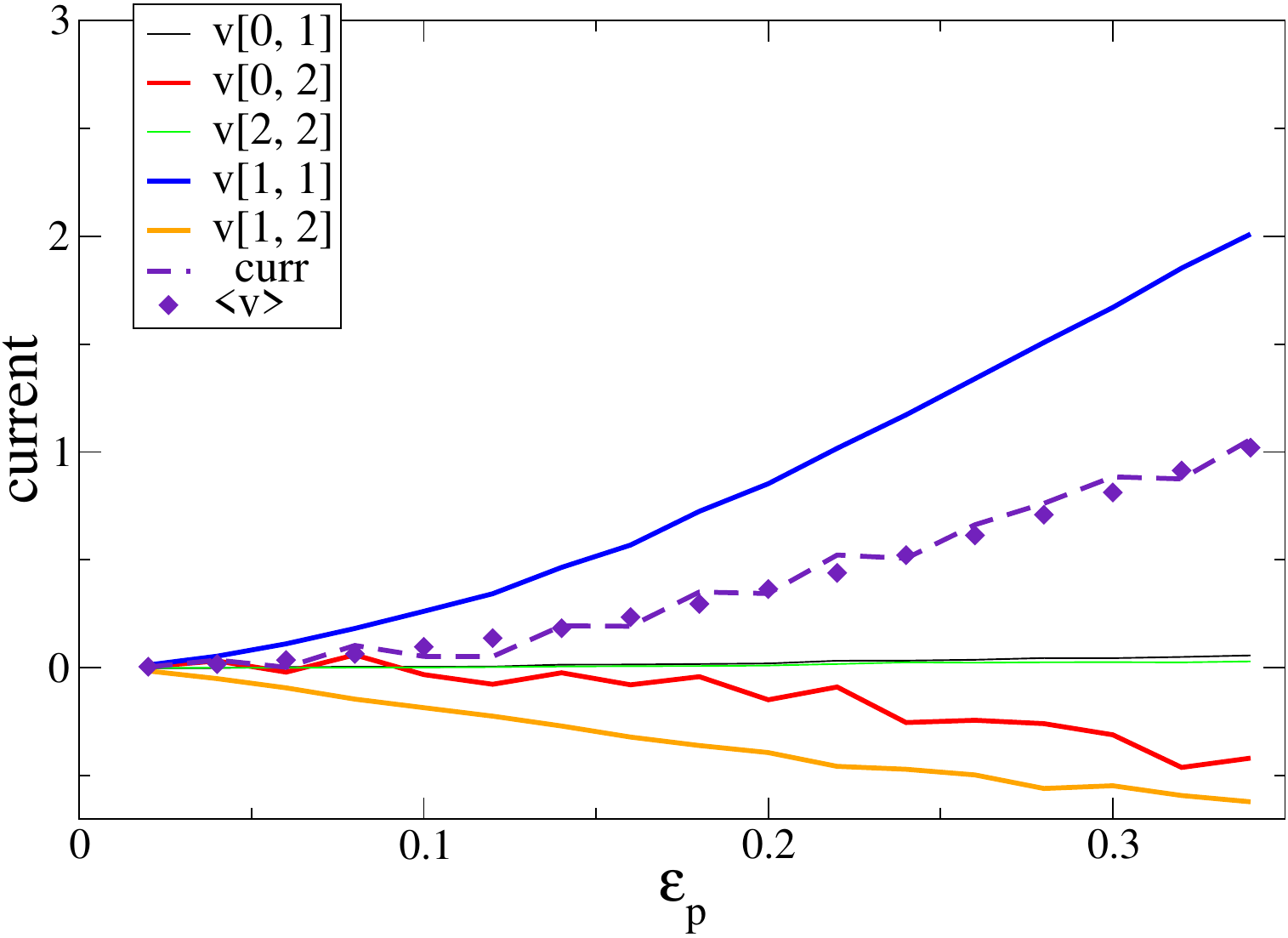}
\caption{
Current and mode contributions to the current as a function of the probe amplitude $\varepsilon_p$ for a system 1D lin$\perp$lin under a probe perturbation propagating to the right (\ref{eq:upm1D:v1b}). Each mode $(l,n)$ has a frequency $\omega=l\delta_p$ and wave number $k=n k_0$. Mode amplitudes are measured in the simulation via (\ref{eq:pfourier:def3}), and their precise contribution to the current determined using  (\ref{eq:opcionB_kp_ne_k0}), taking averages over $1.5\cdot 10^6$ trajectories.
 Units are defined such that $m_a=\hbar=k_l=1$. Here $U_0=50$ and $\Gamma'=7.5$, $\delta_p=10$, and $\phi_p=-\pi/2$. The dashed line is the sum of all current contributions, and the filled diamonds is the current accurately calculated in the simulation from its definition (\ref{eq:currdef}).
\label{fig:fig_v1b_vs_EP}
}
\end{figure}

\subsection{System 1D lin$\perp$lin}

By following the steps presented in Sec.~\ref{sec:sketch}, we obtain the following expansion of the current in terms of mode amplitudes
\begin{eqnarray}
\langle v \rangle_\mathrm{1D} = \frac{m_a}{m_a F_0 g_1 - 2 D_0 k_0} \Bigg[   
-\frac{\text{Im}\left[P_+[0,1]\right] 8 F_0 g_0^2}{k_0}  \nonumber \\
+\frac{\text{Im}\left[P_+[0,2]\right] F_0 \left(-4 g_0 g_1+F_0 k_0/m_a\right)}{k_0}  \nonumber \\
+\frac{\text{Im}\left[e^{i\phi_p}P_+[1,2]\right] 2 F_0 F_p }{m_a}  \nonumber \\
+\frac{ \text{Im}\left[e^{i\phi_p}P_+[1,1]\right] 2 F_p  \delta_p^2}{k_0} \nonumber \\
+\frac{\text{Im}\left[e^{i2\phi_p}P_+[2,2]\right] F_p^2 }{m_a}  
\Bigg]. \label{eq:opcionB_kp_ne_k0}
\end{eqnarray}
Note the only contribution that depends explicitly on the probe frequency $\delta_p$ is the Brillouin mode $(l,n)=(1,1)$, the same propagating mode as the probe potential, all other modes depend explicitly only on the optical potential force $F_0$, probe amplitude $F_p$, or transition rates $g_0$ and $g_1$.  

All terms in (\ref{eq:opcionB_kp_ne_k0}) share a common denominator, $m_a F_0 g_1 - 2 D_0 k_0$, which produce a singularity when is zero, and it will be discussed in Sec.~\ref{sec:inf}.

The expression (\ref{eq:opcionB_kp_ne_k0}) is validated numerically in Fig.~\ref{fig:fig_v1b_vs_EP}, where we depict the contribution of each mode, $v[l,n]$, after numerically calculating the mode amplitudes $P_+[l,n]$ in the simulations using (\ref{eq:pfourier:def3}). As a test, we also show the sum of all contributions and compare it with the current calculated directly in the simulation from its definition  (\ref{eq:currdef}), confirming the analytical calculation. Reduced units are defined,  in all simulations reported in thus paper, such that $m_a=\hbar=k_l=1$.

Previous research \cite{gryn96,renzoni02a,renzoni02b,renzoni03} has focused on one propagating mode, the mode $(l,n)=(1,1)$, which travels at speed $\delta_p/k_0$ and is obviously excited by the probe potential, having the same frequency and wave number. Here this mode is confirmed to provide the most important contribution to the current. However, other modes are also excited, and they even produce a noticeable contribution, as  illustrated  in Fig.~\ref{fig:fig_v1b_vs_EP}. The atomic mode $(1,2)$, which is also propagating to the right, though at the smaller speed $\delta_p/(2k_0)$, as well as the non-propagating mode $(0,2)$, produce contributions to the current in the opposite direction. The total current is, however, dominated by the mode $(1,1)$ and goes in the same direction as the propagating probe potential.

\begin{figure}[t]
\includegraphics[width=8cm]{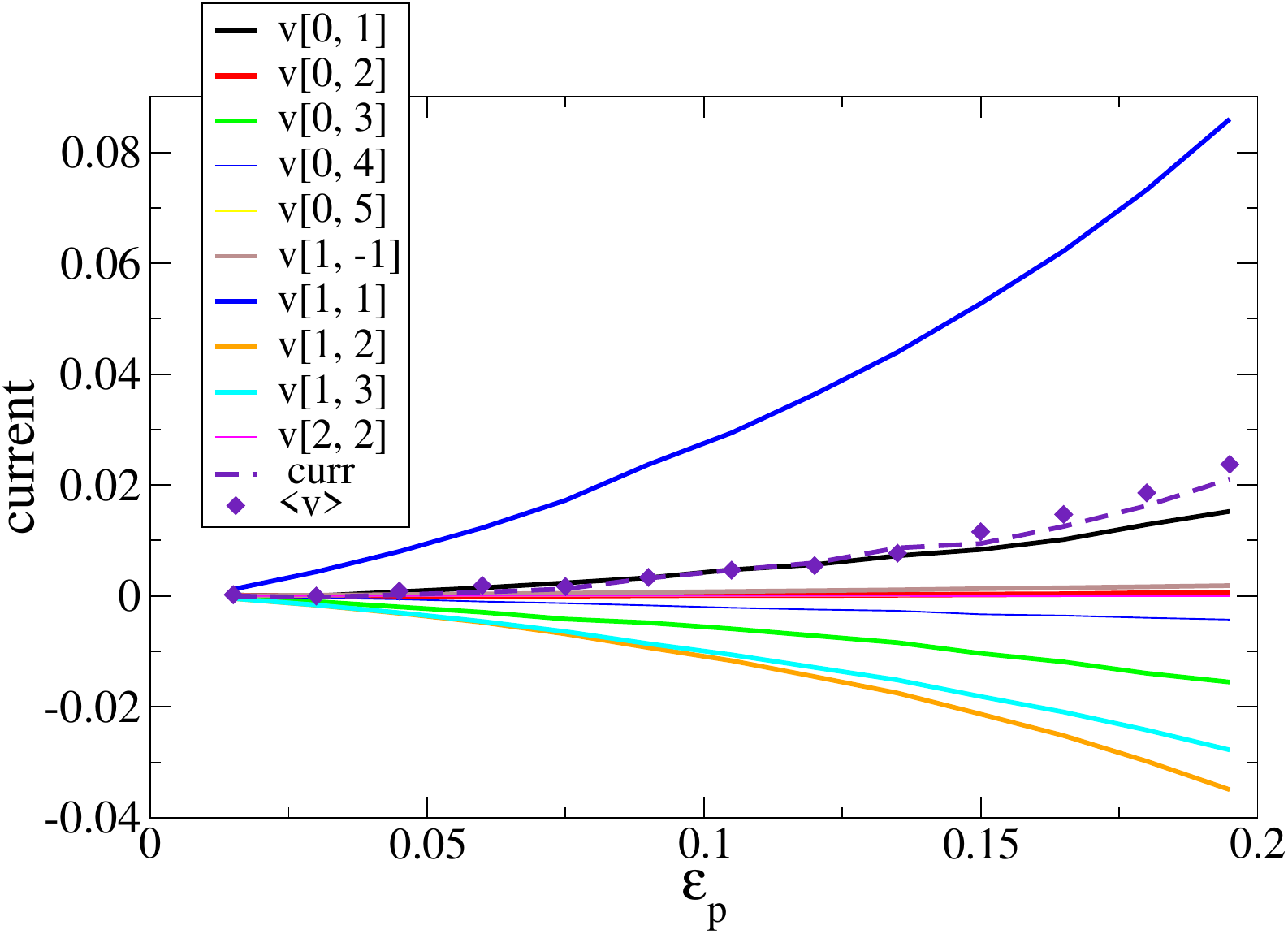}
\caption{
Current and mode contributions to the current as a function of the probe amplitude $\varepsilon_p$ for a system 3D lin$\perp$lin under a probe perturbation propagating to the right in the $x$-direction  (\ref{eq:Upm:3D}), with $U_0=200$, $\Gamma_s=2.85$, $\theta_x=\theta_y=25$º, $\delta_p=12$, and $\phi_p=\pi$. The dashed line is the sum of all current contributions, and the filled diamonds is the current measured directly in the simulation.
\label{fig:fig_v1b_vs_EP}
}
\end{figure}

\subsection{System 3D-lin$\perp$lin}
In the system 3D--lin$\perp$lin, the additional terms (\ref{eq:F0:2})--(\ref{eq:g0:2}) produce more mode contributions to the current, appearing both more contributing terms for the previously discussed modes, and new contributions from other modes.
\begin{eqnarray}
\langle v \rangle_\mathrm{3D} = \frac{m_a}{m_a F_0 g_1 - 2 D_0 k_0} \Bigg[   \nonumber \\
-\frac{\text{Im}\left[P_+[0,1,0]\right] F_0\left( 8 g_0^2 {\color{blue}-4g_2^2/3+F_0 k_0/(2m_a)}  \right)  }{k_0}  \nonumber \\
+\frac{\text{Im}\left[P_+[0,2,0]\right] F_0 \left(-4 g_0 g_1 {\color{blue}-8g_1 g_2/3} +F_0 k_0/m_a\right)}{k_0}  \nonumber \\
\color{blue}+\frac{\text{Im}\left[P_+[0,3,0]\right] F_0 \left(-16 g_0 g_1/3 - 2 g_2^2 -3F_0 k_0/(2m_a)\right)}{k_0}  \nonumber \\
\color{blue}+\frac{\text{Im}\left[P_+[0,4,0]\right] F_0 \left(-2 g_1 g_2/3 +F_0 k_0/(2m_a)\right)}{k_0}  \nonumber \\
\color{blue}-\frac{\text{Im}\left[P_+[0,5,0]\right] F_0  2 g_2^2 }{3 k_0} 
+\frac{\text{Im}\left[e^{i\phi_p}P_+[1,1,1]\right] 2 F_0 F_p }{m_a}  \nonumber \\
\color{blue} -\frac{\text{Re}\left[e^{i\phi_p}P_+[1,-2,1]\right] F_0 F_p }{2 m_a} 
\color{blue} +\frac{\text{Re}\left[e^{i\phi_p}P_+[1,2,1]\right] 3 F_0 F_p }{2 m_a} \nonumber \\
+\frac{ \text{Im}\left[e^{i\phi_p}P_+[1,0,1]\right] 2 F_p  \delta_p^2}{k_0} 
+\frac{\text{Im}\left[e^{i2\phi_p}P_+[2,0,2]\right] F_p^2 }{m_a}  
\Bigg]. \label{eq:3Dlin}  \nonumber \\
\end{eqnarray}

Like in the previous system, the biggest contributor to the current is the propagating mode $(1,1)$, producing movement in the same direction as the propagating probe potential, here with the help of  non-propagating mode $(0,1)$, which is also present in the system 1D lin$\perp$lin ---also producing there a current in the same direction, but with a much less appreciable contribution. 
New contributions from the non-propagating modes $(0,3)$, $(0,4)$ and $(0,5)$---due to the extra mode-connecting terms in (\ref{eq:Upm:3D}) and (\ref{eq:gammapm})---are observed, all pointing to the opposite direction of propagation. A similar feature is observed in the new modes $(1,-1)$ and $(1,3)$, which together with the propagating mode $(1,2)$ produce current in the opposite direction. The overall current is much smaller---about four times smaller---than the one produced alone by the Brillouin mode $(1,1)$ due to those counter-propagating contributions.

\section{Into the regime of infinite density}
\label{sec:inf}
The fact that the optical potential force  $\mathcal{F}_{\pm1}^{(0)\pm}$, Eq.~(\ref{eq:F0:1}), inverts its sign when the atomic state is changed is crucially coupled to the similar state-alternating sign in the transition rate  $\gamma_{\pm1}^\pm$, Eq.~(\ref{eq:g0:1}), being responsible for the celebrated Sisyphus cooling---a kind of dissipative mechanism due to the transitions taking place with higher probability  when the atoms are at the top of the optical potential barriers, thus the atom spends most of the time climbing hills, like the punishment inflicted in the king Sisyphus of ancient Greek mythology. 
However, it is known the potential should not be too shallow \cite{cohentan92}, or the dynamics may exploit in the form of superdiffusion \cite{wichol12} and non-normalizable probability densities \cite{lutren13,hol15}.  Indeed, the calculations presented in Sec. \ref{sec:newtheor} indicate that the current diverges to infinity, in both studied systems, when  
\begin{equation}
F_0 g_1-2D_0 k_0/m_a = 0,
\label{eq:curr:inf}
\end{equation}
because the left hand side of (\ref{eq:curr:inf}) appears as a common factor in the denominator  in the formulas for the current,  (\ref{eq:opcionB_kp_ne_k0}) and (\ref{eq:3Dlin}). It is a consequence of the appearance of a term proportional to $\partial \mathcal{P}^+_{0,0,0}/\partial q +\partial \mathcal{P}^-_{0,0,0}/\partial q$   in the right hand side of (\ref{eq:curr:sol}) (with $n=1$), which has to be taken to the left hand side in order to complete the calculation. 

In fact, the same thing happens in any moment $\langle v^n\rangle$.  Using (\ref{eq:moment}) and  (\ref{eq:curr:sol}), and looking for the terms in the right hand side that yield a contribution containing $\partial^n \mathcal{P}^+_{0,0,0}/\partial q^n +\partial^n \mathcal{P}^-_{0,0,0}/\partial q^n$, i.e. the terms proportional to $\mathcal{F}^{(0)\pm}_{+1}\gamma^\pm_{-1}$ and $\mathcal{F}^{(0)\pm}_{-1}\gamma^\pm_{+1}$, 
 yields the common factor, $[1-F_0 g_1 m_a/(D_0 k_0(n+1)]^{-1}$, and thus the following singularity condition,
\begin{equation}
F_0 g_1-(1+n)D_0 k_0/m_a = 0,
\label{eq:moment:inf}
\end{equation}
which in terms of the potential depth is written as
\begin{equation}
 U_0 = \frac{2(1+n) D_0}{g_1 m_a}.
 \label{eq:U0:inf}
 \end{equation}
Though this result has been obtained from the semiclassical equations (\ref{eq:fp}), which do not explicitly contain all the diffusive terms predicted by the semiclassical derivation, that is, the terms $\partial^2 D_{\mp\pm} P_{\mp}/\partial p^2$ and the spatial dependence of the noise coefficient, the same analysis in the more complex system model shows that the exact same result (\ref{eq:moment:inf}) is obtained if one considers $D_0$ as the spatial average of all diffussive terms. 

This singularity can be identified with the threshold for the transition into an anomalous regime, in which the $n$ moment of the velocity distribution, $\langle v(t)^n\rangle$, diverges for long times. The result (\ref{eq:U0:inf}) is in perfect agreement with the prediction \cite{kessler10,hol15} for the system 1D lin$\perp$lin obtained from a simplified, approximate Fokker-Planck equation \cite{cohentan92} for shallow lattices, when  the noise stregth $D_0$ contains the spatial average of both diffusive terms $D_{\mp\pm}$ and $D_{\pm\pm}$ \cite{petsas99}, i.e. $D_0=(35+6)\hbar^2 k_l^2\Gamma'/90$. This is remarkable, as our method uses the full semiclassical equations (\ref{eq:fp}), and hence, unlike the approximate Fokker-Planck equation, it does explicitly takes into account the microscopic origin of Sisyphus cooling, including the periodicity of the lattice potential and transitions between the atomic sublevels. 

The only assumption made by the current method is the existence of the Taylor expansion (\ref{eq:taylor2}), which is numerically confirmed by the simulations presented in the paper in the regime above the singularity. In contrast, in shallow lattices---i.e. with potential wells $U_0$ below the threshold given by (\ref{eq:U0:inf})---this assumption does not hold, giving rise to the regime of infinite densities,  with diverging moments and non-normalizable densities. 

In the system 3D-lin$\perp$lin,  to accurately cover the shallow potential regime, $D_0$ should also include the spatial average the diffusive term $D_{\mp\pm}$, thus $D_0=(1+1/5)5\hbar^2k_0^2\Gamma_s/18$.


\section{Conclusions}
\label{sec:con}
Starting from the semiclassical equations for the atomic phase densities, we have presented an exact method to calculate the precise contribution to the current of the excited atomic density waves in cold atoms under dissipative optical lattices. Explicit analyitcal expressions are provided for the popular system 1D lin$\perp$lin under a simple---but realizable---symmetry breaking perturbation, as well as the one-dimensional model that results from focusing in a specific direction in the 3D-lin$\perp$lin system.  

The analytical results are validated with numerical simulations of the stochastic atomic trajectories associated with the semiclassical equations. They show that several atomic modes, not just the one associated with the perturbation, contribute relevantly to the directed motion.  

 Additionally, the analytical solution predicts a singularity for each moment of the velocity distribution, which is identified with the threshold for the transition to the regime of infinite density. The threshold values of the present work are identical to the ones obtained from an approximated Fokker-Planck equation of the semiclassical equations, derived by spatial and Zeeman level averaging the latter, in the system 1D lin$\perp$lin, thus providing a solid ground to previous analytical results. 
 
 Finally, it is worth mentioning that the proposed method can easily be applied to more complex systems and driving perturbations.





\begin{acknowledgments}
DC acknowledges financial support from the Ministerio de Econom\'ia y Competitividad of Spain, Grant No. PID2019-105316GB-I00.
\end{acknowledgments}


 


\begin{thebibliography}{29}%
\makeatletter
\providecommand \@ifxundefined [1]{%
 \@ifx{#1\undefined}
}%
\providecommand \@ifnum [1]{%
 \ifnum #1\expandafter \@firstoftwo
 \else \expandafter \@secondoftwo
 \fi
}%
\providecommand \@ifx [1]{%
 \ifx #1\expandafter \@firstoftwo
 \else \expandafter \@secondoftwo
 \fi
}%
\providecommand \natexlab [1]{#1}%
\providecommand \enquote  [1]{``#1''}%
\providecommand \bibnamefont  [1]{#1}%
\providecommand \bibfnamefont [1]{#1}%
\providecommand \citenamefont [1]{#1}%
\providecommand \href@noop [0]{\@secondoftwo}%
\providecommand \href [0]{\begingroup \@sanitize@url \@href}%
\providecommand \@href[1]{\@@startlink{#1}\@@href}%
\providecommand \@@href[1]{\endgroup#1\@@endlink}%
\providecommand \@sanitize@url [0]{\catcode `\\12\catcode `\$12\catcode
  `\&12\catcode `\#12\catcode `\^12\catcode `\_12\catcode `\%12\relax}%
\providecommand \@@startlink[1]{}%
\providecommand \@@endlink[0]{}%
\providecommand \url  [0]{\begingroup\@sanitize@url \@url }%
\providecommand \@url [1]{\endgroup\@href {#1}{\urlprefix }}%
\providecommand \urlprefix  [0]{URL }%
\providecommand \Eprint [0]{\href }%
\providecommand \doibase [0]{http://dx.doi.org/}%
\providecommand \selectlanguage [0]{\@gobble}%
\providecommand \bibinfo  [0]{\@secondoftwo}%
\providecommand \bibfield  [0]{\@secondoftwo}%
\providecommand \translation [1]{[#1]}%
\providecommand \BibitemOpen [0]{}%
\providecommand \bibitemStop [0]{}%
\providecommand \bibitemNoStop [0]{.\EOS\space}%
\providecommand \EOS [0]{\spacefactor3000\relax}%
\providecommand \BibitemShut  [1]{\csname bibitem#1\endcsname}%
\let\auto@bib@innerbib\@empty
\bibitem [{\citenamefont {Schreck}\ and\ \citenamefont {van
  Druten}(2021)}]{schreck21}%
  \BibitemOpen
  \bibfield  {author} {\bibinfo {author} {\bibfnamefont {F.}~\bibnamefont
  {Schreck}}\ and\ \bibinfo {author} {\bibfnamefont {K.}~\bibnamefont {van
  Druten}},\ }\bibfield  {title} {\enquote {\bibinfo {title} {Laser cooling for
  quantum gases},}\ }\href@noop {} {\bibfield  {journal} {\bibinfo  {journal}
  {Nat. Phys.}\ }\textbf {\bibinfo {volume} {17}},\ \bibinfo {pages} {1296}
  (\bibinfo {year} {2021})}\BibitemShut {NoStop}%
\bibitem [{\citenamefont {Grynberg}\ and\ \citenamefont
  {Robilliard}(2001{\natexlab{a}})}]{grynreview}%
  \BibitemOpen
  \bibfield  {author} {\bibinfo {author} {\bibfnamefont {G.}~\bibnamefont
  {Grynberg}}\ and\ \bibinfo {author} {\bibfnamefont {C.}~\bibnamefont
  {Robilliard}},\ }\enquote {\bibinfo {title} {Cold atoms in dissipative
  optical lattices},}\ \ (\bibinfo {year} {2001})\ p.\ \bibinfo {pages}
  {335}\BibitemShut {NoStop}%
\bibitem [{\citenamefont {Mennerat-Robilliard}\ \emph
  {et~al.}(1999)\citenamefont {Mennerat-Robilliard}, \citenamefont {Lucas},
  \citenamefont {Guibal}, \citenamefont {Tabosa}, \citenamefont {Jurczak},
  \citenamefont {Courtois},\ and\ \citenamefont {Grynberg}}]{cecile99}%
  \BibitemOpen
  \bibfield  {author} {\bibinfo {author} {\bibfnamefont {C.}~\bibnamefont
  {Mennerat-Robilliard}}, \bibinfo {author} {\bibfnamefont {D.}~\bibnamefont
  {Lucas}}, \bibinfo {author} {\bibfnamefont {S.}~\bibnamefont {Guibal}},
  \bibinfo {author} {\bibfnamefont {J.}~\bibnamefont {Tabosa}}, \bibinfo
  {author} {\bibfnamefont {C.}~\bibnamefont {Jurczak}}, \bibinfo {author}
  {\bibfnamefont {J.-Y.}\ \bibnamefont {Courtois}}, \ and\ \bibinfo {author}
  {\bibfnamefont {G.}~\bibnamefont {Grynberg}},\ }\bibfield  {title} {\enquote
  {\bibinfo {title} {Ratchet for cold rubidium atoms: The asymmetric optical
  lattice},}\ }\href@noop {} {\bibfield  {journal} {\bibinfo  {journal} {Phys.
  Rev. Lett.}\ }\textbf {\bibinfo {volume} {82}},\ \bibinfo {pages} {851}
  (\bibinfo {year} {1999})}\BibitemShut {NoStop}%
\bibitem [{\citenamefont {Schiavoni}\ \emph {et~al.}(2003)\citenamefont
  {Schiavoni}, \citenamefont {Sanchez-Palencia}, \citenamefont {Renzoni},\ and\
  \citenamefont {Grynberg}}]{schiavoni2003}%
  \BibitemOpen
  \bibfield  {author} {\bibinfo {author} {\bibfnamefont {M.}~\bibnamefont
  {Schiavoni}}, \bibinfo {author} {\bibfnamefont {L.}~\bibnamefont
  {Sanchez-Palencia}}, \bibinfo {author} {\bibfnamefont {F.}~\bibnamefont
  {Renzoni}}, \ and\ \bibinfo {author} {\bibfnamefont {G.}~\bibnamefont
  {Grynberg}},\ }\bibfield  {title} {\enquote {\bibinfo {title} {Phase control
  of directed diffusion in a symmetric optical lattice},}\ }\href@noop {}
  {\bibfield  {journal} {\bibinfo  {journal} {Phys. Rev. Lett.}\ }\textbf
  {\bibinfo {volume} {90}},\ \bibinfo {pages} {094101} (\bibinfo {year}
  {2003})}\BibitemShut {NoStop}%
\bibitem [{\citenamefont {Gommers}\ \emph {et~al.}(2005)\citenamefont
  {Gommers}, \citenamefont {Bergamini},\ and\ \citenamefont
  {Renzoni}}]{gommers2005}%
  \BibitemOpen
  \bibfield  {author} {\bibinfo {author} {\bibfnamefont {R.}~\bibnamefont
  {Gommers}}, \bibinfo {author} {\bibfnamefont {S.}~\bibnamefont {Bergamini}},
  \ and\ \bibinfo {author} {\bibfnamefont {F.}~\bibnamefont {Renzoni}},\
  }\bibfield  {title} {\enquote {\bibinfo {title} {Dissipation-induced symmetry
  breaking in a driven optical lattice},}\ }\href@noop {} {\bibfield  {journal}
  {\bibinfo  {journal} {Phys. Rev. Lett.}\ }\textbf {\bibinfo {volume} {95}},\
  \bibinfo {pages} {073003} (\bibinfo {year} {2005})}\BibitemShut {NoStop}%
\bibitem [{\citenamefont {Gommers}\ \emph {et~al.}(2006)\citenamefont
  {Gommers}, \citenamefont {Denisov},\ and\ \citenamefont
  {Renzoni}}]{gommers2006}%
  \BibitemOpen
  \bibfield  {author} {\bibinfo {author} {\bibfnamefont {R.}~\bibnamefont
  {Gommers}}, \bibinfo {author} {\bibfnamefont {S.}~\bibnamefont {Denisov}}, \
  and\ \bibinfo {author} {\bibfnamefont {F.}~\bibnamefont {Renzoni}},\
  }\bibfield  {title} {\enquote {\bibinfo {title} {Quasiperiodically driven
  ratchets for cold atoms},}\ }\href@noop {} {\bibfield  {journal} {\bibinfo
  {journal} {Phys. Rev. Lett.}\ }\textbf {\bibinfo {volume} {96}},\ \bibinfo
  {pages} {240604} (\bibinfo {year} {2006})}\BibitemShut {NoStop}%
\bibitem [{\citenamefont {Jones}\ \emph {et~al.}(2007)\citenamefont {Jones},
  \citenamefont {Goonasekera}, \citenamefont {Meacher}, \citenamefont
  {Jonckheere},\ and\ \citenamefont {Monteiro}}]{jones2007}%
  \BibitemOpen
  \bibfield  {author} {\bibinfo {author} {\bibfnamefont {P.~H.}\ \bibnamefont
  {Jones}}, \bibinfo {author} {\bibfnamefont {M.}~\bibnamefont {Goonasekera}},
  \bibinfo {author} {\bibfnamefont {D.~R.}\ \bibnamefont {Meacher}}, \bibinfo
  {author} {\bibfnamefont {T.}~\bibnamefont {Jonckheere}}, \ and\ \bibinfo
  {author} {\bibfnamefont {T.~S.}\ \bibnamefont {Monteiro}},\ }\bibfield
  {title} {\enquote {\bibinfo {title} {Directed motion for delta-kicked atoms
  with broken symmetries: Comparison between theory and experiment},}\
  }\href@noop {} {\bibfield  {journal} {\bibinfo  {journal} {Phys. Rev. Lett.}\
  }\textbf {\bibinfo {volume} {98}},\ \bibinfo {pages} {073002} (\bibinfo
  {year} {2007})}\BibitemShut {NoStop}%
\bibitem [{\citenamefont {Wickenbrock}\ \emph {et~al.}(2012)\citenamefont
  {Wickenbrock}, \citenamefont {Holz}, \citenamefont {Wahab}, \citenamefont
  {Phoonthong}, \citenamefont {Cubero},\ and\ \citenamefont
  {Renzoni}}]{wichol12}%
  \BibitemOpen
  \bibfield  {author} {\bibinfo {author} {\bibfnamefont {A.}~\bibnamefont
  {Wickenbrock}}, \bibinfo {author} {\bibfnamefont {P.~C.}\ \bibnamefont
  {Holz}}, \bibinfo {author} {\bibfnamefont {N.~A.~Abdul}\ \bibnamefont
  {Wahab}}, \bibinfo {author} {\bibfnamefont {P.}~\bibnamefont {Phoonthong}},
  \bibinfo {author} {\bibfnamefont {D.}~\bibnamefont {Cubero}}, \ and\ \bibinfo
  {author} {\bibfnamefont {F.}~\bibnamefont {Renzoni}},\ }\bibfield  {title}
  {\enquote {\bibinfo {title} {Vibrational mechanics in an optical lattice:
  Controlling transport via potential renormalization},}\ }\href@noop {}
  {\bibfield  {journal} {\bibinfo  {journal} {Phys. Rev. Lett.}\ }\textbf
  {\bibinfo {volume} {108}},\ \bibinfo {pages} {020603} (\bibinfo {year}
  {2012})}\BibitemShut {NoStop}%
\bibitem [{\citenamefont {Salger}\ \emph {et~al.}(2009)\citenamefont {Salger},
  \citenamefont {Kling}, \citenamefont {Hecking}, \citenamefont {Geckeler},
  \citenamefont {Morales-Molina},\ and\ \citenamefont {Weitz}}]{salger2009}%
  \BibitemOpen
  \bibfield  {author} {\bibinfo {author} {\bibfnamefont {T.}~\bibnamefont
  {Salger}}, \bibinfo {author} {\bibfnamefont {S.}~\bibnamefont {Kling}},
  \bibinfo {author} {\bibfnamefont {T.}~\bibnamefont {Hecking}}, \bibinfo
  {author} {\bibfnamefont {C.}~\bibnamefont {Geckeler}}, \bibinfo {author}
  {\bibfnamefont {L.}~\bibnamefont {Morales-Molina}}, \ and\ \bibinfo {author}
  {\bibfnamefont {M.}~\bibnamefont {Weitz}},\ }\bibfield  {title} {\enquote
  {\bibinfo {title} {Directed transport of atoms in a hamiltonian quantum
  ratchet},}\ }\href@noop {} {\bibfield  {journal} {\bibinfo  {journal}
  {Science}\ }\textbf {\bibinfo {volume} {326}},\ \bibinfo {pages} {1241}
  (\bibinfo {year} {2009})}\BibitemShut {NoStop}%
\bibitem [{\citenamefont {Grossert}\ \emph {et~al.}(2016)\citenamefont
  {Grossert}, \citenamefont {Leder}, \citenamefont {Denisov}, \citenamefont
  {H\"anggi},\ and\ \citenamefont {Weitz}}]{denisov15}%
  \BibitemOpen
  \bibfield  {author} {\bibinfo {author} {\bibfnamefont {C.}~\bibnamefont
  {Grossert}}, \bibinfo {author} {\bibfnamefont {M.}~\bibnamefont {Leder}},
  \bibinfo {author} {\bibfnamefont {S.}~\bibnamefont {Denisov}}, \bibinfo
  {author} {\bibfnamefont {P.}~\bibnamefont {H\"anggi}}, \ and\ \bibinfo
  {author} {\bibfnamefont {M.}~\bibnamefont {Weitz}},\ }\bibfield  {title}
  {\enquote {\bibinfo {title} {Experimental control of transport resonances in
  a coherent quantum rocking ratchet},}\ }\href@noop {} {\bibfield  {journal}
  {\bibinfo  {journal} {Nat. Commun.}\ }\textbf {\bibinfo {volume} {7}},\
  \bibinfo {pages} {10440} (\bibinfo {year} {2016})}\BibitemShut {NoStop}%
\bibitem [{\citenamefont {H\"anggi}\ and\ \citenamefont
  {Marchesoni}(2009)}]{hanmar09}%
  \BibitemOpen
  \bibfield  {author} {\bibinfo {author} {\bibfnamefont {P.}~\bibnamefont
  {H\"anggi}}\ and\ \bibinfo {author} {\bibfnamefont {F.}~\bibnamefont
  {Marchesoni}},\ }\bibfield  {title} {\enquote {\bibinfo {title} {Artificial
  brownian motors: Controlling transport on the nanoscale},}\ }\href@noop {}
  {\bibfield  {journal} {\bibinfo  {journal} {Rev. Mod. Phys.}\ }\textbf
  {\bibinfo {volume} {81}},\ \bibinfo {pages} {387} (\bibinfo {year}
  {2009})}\BibitemShut {NoStop}%
\bibitem [{\citenamefont {Cubero}\ and\ \citenamefont
  {Renzoni}(2016{\natexlab{a}})}]{cubren16}%
  \BibitemOpen
  \bibfield  {author} {\bibinfo {author} {\bibfnamefont {D.}~\bibnamefont
  {Cubero}}\ and\ \bibinfo {author} {\bibfnamefont {F.}~\bibnamefont
  {Renzoni}},\ }\href@noop {} {\emph {\bibinfo {title} {Brownian Ratchets: From
  Statistical Physics to Bio and Nano-motors}}}\ (\bibinfo  {publisher}
  {Cambridge University Press},\ \bibinfo {address} {Cambridge},\ \bibinfo
  {year} {2016})\BibitemShut {NoStop}%
\bibitem [{\citenamefont {Lutz}\ and\ \citenamefont
  {Renzoni}(2013)}]{lutren13}%
  \BibitemOpen
  \bibfield  {author} {\bibinfo {author} {\bibfnamefont {E.}~\bibnamefont
  {Lutz}}\ and\ \bibinfo {author} {\bibfnamefont {F.}~\bibnamefont {Renzoni}},\
  }\bibfield  {title} {\enquote {\bibinfo {title} {Beyond boltzmann–gibbs
  statistical mechanics in optical lattices},}\ }\href@noop {} {\bibfield
  {journal} {\bibinfo  {journal} {Nature Physics}\ }\textbf {\bibinfo {volume}
  {9}},\ \bibinfo {pages} {615} (\bibinfo {year} {2013})}\BibitemShut {NoStop}%
\bibitem [{\citenamefont {Kessler}\ and\ \citenamefont
  {Barkai}(2010)}]{kessler10}%
  \BibitemOpen
  \bibfield  {author} {\bibinfo {author} {\bibfnamefont {D.~A.}\ \bibnamefont
  {Kessler}}\ and\ \bibinfo {author} {\bibfnamefont {E.}~\bibnamefont
  {Barkai}},\ }\bibfield  {title} {\enquote {\bibinfo {title} {Infinite
  covariant density for diffusion in logarithmic potentials and optical
  lattices},}\ }\href@noop {} {\bibfield  {journal} {\bibinfo  {journal} {Phys.
  Rev. Lett.}\ }\textbf {\bibinfo {volume} {105}},\ \bibinfo {pages} {120602}
  (\bibinfo {year} {2010})}\BibitemShut {NoStop}%
\bibitem [{\citenamefont {Holz}\ \emph {et~al.}(2015)\citenamefont {Holz},
  \citenamefont {Dechant},\ and\ \citenamefont {Lutz}}]{hol15}%
  \BibitemOpen
  \bibfield  {author} {\bibinfo {author} {\bibfnamefont {P.C.}\ \bibnamefont
  {Holz}}, \bibinfo {author} {\bibfnamefont {A.}~\bibnamefont {Dechant}}, \
  and\ \bibinfo {author} {\bibfnamefont {E.}~\bibnamefont {Lutz}},\ }\bibfield
  {title} {\enquote {\bibinfo {title} {Infinite density for cold atoms in
  shallow optical lattices},}\ }\href@noop {} {\bibfield  {journal} {\bibinfo
  {journal} {EPL}\ }\textbf {\bibinfo {volume} {109}},\ \bibinfo {pages}
  {23001} (\bibinfo {year} {2015})}\BibitemShut {NoStop}%
\bibitem [{\citenamefont {Barkai}\ \emph {et~al.}(2021)\citenamefont {Barkai},
  \citenamefont {Radons}, \citenamefont {Akimoto}, \citenamefont {Schreck},\
  and\ \citenamefont {van Druten}}]{barkai21}%
  \BibitemOpen
  \bibfield  {author} {\bibinfo {author} {\bibfnamefont {E.}~\bibnamefont
  {Barkai}}, \bibinfo {author} {\bibfnamefont {G.}~\bibnamefont {Radons}},
  \bibinfo {author} {\bibfnamefont {T.}~\bibnamefont {Akimoto}}, \bibinfo
  {author} {\bibfnamefont {F.}~\bibnamefont {Schreck}}, \ and\ \bibinfo
  {author} {\bibfnamefont {K.}~\bibnamefont {van Druten}},\ }\bibfield  {title}
  {\enquote {\bibinfo {title} {Transitions in the ergodicity of
  subrecoil-laser-cooled gases},}\ }\href@noop {} {\bibfield  {journal}
  {\bibinfo  {journal} {Phys. Rev. Lett.}\ }\textbf {\bibinfo {volume} {127}},\
  \bibinfo {pages} {140605} (\bibinfo {year} {2021})}\BibitemShut {NoStop}%
\bibitem [{\citenamefont {Akimoto}\ \emph {et~al.}(2022)\citenamefont
  {Akimoto}, \citenamefont {Barkai},\ and\ \citenamefont {Radons}}]{barkai22}%
  \BibitemOpen
  \bibfield  {author} {\bibinfo {author} {\bibfnamefont {T.}~\bibnamefont
  {Akimoto}}, \bibinfo {author} {\bibfnamefont {E.}~\bibnamefont {Barkai}}, \
  and\ \bibinfo {author} {\bibfnamefont {G.}~\bibnamefont {Radons}},\
  }\bibfield  {title} {\enquote {\bibinfo {title} {Infinite ergodic theory for
  three heterogeneous stochastic models with application to subrecoil laser
  cooling},}\ }\href@noop {} {\bibfield  {journal} {\bibinfo  {journal} {Phys.
  Rev. E}\ }\textbf {\bibinfo {volume} {105}},\ \bibinfo {pages} {064126}
  (\bibinfo {year} {2022})}\BibitemShut {NoStop}%
\bibitem [{\citenamefont {Courtois}\ \emph {et~al.}(1996)\citenamefont
  {Courtois}, \citenamefont {Guibal}, \citenamefont {Meacher}, \citenamefont
  {Verkerk},\ and\ \citenamefont {Grynberg}}]{gryn96}%
  \BibitemOpen
  \bibfield  {author} {\bibinfo {author} {\bibfnamefont {J.~Y.}\ \bibnamefont
  {Courtois}}, \bibinfo {author} {\bibfnamefont {S.}~\bibnamefont {Guibal}},
  \bibinfo {author} {\bibfnamefont {D.~R.}\ \bibnamefont {Meacher}}, \bibinfo
  {author} {\bibfnamefont {P.}~\bibnamefont {Verkerk}}, \ and\ \bibinfo
  {author} {\bibfnamefont {G.}~\bibnamefont {Grynberg}},\ }\bibfield  {title}
  {\enquote {\bibinfo {title} {Propagating elementary excitation in a dilute
  optical lattice},}\ }\href@noop {} {\bibfield  {journal} {\bibinfo  {journal}
  {Phys. Rev. Lett.}\ }\textbf {\bibinfo {volume} {77}},\ \bibinfo {pages} {40}
  (\bibinfo {year} {1996})}\BibitemShut {NoStop}%
\bibitem [{\citenamefont {Sanchez-Palencia}\ \emph {et~al.}(2002)\citenamefont
  {Sanchez-Palencia}, \citenamefont {Carminati}, \citenamefont {Schiavoni},
  \citenamefont {Renzoni},\ and\ \citenamefont {Grynberg}}]{renzoni02a}%
  \BibitemOpen
  \bibfield  {author} {\bibinfo {author} {\bibfnamefont {L.}~\bibnamefont
  {Sanchez-Palencia}}, \bibinfo {author} {\bibfnamefont {F.-R.}\ \bibnamefont
  {Carminati}}, \bibinfo {author} {\bibfnamefont {M.}~\bibnamefont
  {Schiavoni}}, \bibinfo {author} {\bibfnamefont {F.}~\bibnamefont {Renzoni}},
  \ and\ \bibinfo {author} {\bibfnamefont {G.}~\bibnamefont {Grynberg}},\
  }\bibfield  {title} {\enquote {\bibinfo {title} {Brillouin propagation modes
  in optical lattices: Interpretation in terms of nonconventional stochastic
  resonance},}\ }\href@noop {} {\bibfield  {journal} {\bibinfo  {journal}
  {Phys. Rev. Lett.}\ }\textbf {\bibinfo {volume} {88}},\ \bibinfo {pages}
  {133903--1} (\bibinfo {year} {2002})}\BibitemShut {NoStop}%
\bibitem [{\citenamefont {Borromeo}\ and\ \citenamefont
  {Marchesoni}(2007)}]{fabio07}%
  \BibitemOpen
  \bibfield  {author} {\bibinfo {author} {\bibfnamefont {M.}~\bibnamefont
  {Borromeo}}\ and\ \bibinfo {author} {\bibfnamefont {F.}~\bibnamefont
  {Marchesoni}},\ }\bibfield  {title} {\enquote {\bibinfo {title} {Artificial
  sieves for quasimassless particles},}\ }\href@noop {} {\bibfield  {journal}
  {\bibinfo  {journal} {Phys. Rev. Lett.}\ }\textbf {\bibinfo {volume} {99}},\
  \bibinfo {pages} {150605} (\bibinfo {year} {2007})}\BibitemShut {NoStop}%
\bibitem [{\citenamefont {Castin}\ \emph {et~al.}(1991)\citenamefont {Castin},
  \citenamefont {Dalibard},\ and\ \citenamefont
  {Cohen.Tannoudji}}]{cohentan92}%
  \BibitemOpen
  \bibfield  {author} {\bibinfo {author} {\bibfnamefont {Y.}~\bibnamefont
  {Castin}}, \bibinfo {author} {\bibfnamefont {J.}~\bibnamefont {Dalibard}}, \
  and\ \bibinfo {author} {\bibfnamefont {C.}~\bibnamefont {Cohen.Tannoudji}},\
  }\enquote {\bibinfo {title} {Light induced kinetic effects on atoms, ions and
  molecules},}\ \ (\bibinfo  {publisher} {ETS Editrice},\ \bibinfo {address}
  {Pisa},\ \bibinfo {year} {1991})\ Chap.\ \bibinfo {chapter} {The limits of
  Sisyphus cooling}, pp.\ \bibinfo {pages} {5--24}\BibitemShut {NoStop}%
\bibitem [{\citenamefont {Grynberg}\ and\ \citenamefont
  {Robilliard}(2001{\natexlab{b}})}]{grynberg2001}%
  \BibitemOpen
  \bibfield  {author} {\bibinfo {author} {\bibfnamefont {G.}~\bibnamefont
  {Grynberg}}\ and\ \bibinfo {author} {\bibfnamefont {C.}~\bibnamefont
  {Robilliard}},\ }\bibfield  {title} {\enquote {\bibinfo {title} {Cold atoms
  in dissipative optical lattices},}\ }\href@noop {} {\bibfield  {journal}
  {\bibinfo  {journal} {Phys. Rep.}\ }\textbf {\bibinfo {volume} {355}},\
  \bibinfo {pages} {335} (\bibinfo {year} {2001}{\natexlab{b}})}\BibitemShut
  {NoStop}%
\bibitem [{\citenamefont {Petsas}\ \emph {et~al.}(1999)\citenamefont {Petsas},
  \citenamefont {Grynberg},\ and\ \citenamefont {Courtois}}]{petsas99}%
  \BibitemOpen
  \bibfield  {author} {\bibinfo {author} {\bibfnamefont {K.}~\bibnamefont
  {Petsas}}, \bibinfo {author} {\bibfnamefont {G.}~\bibnamefont {Grynberg}}, \
  and\ \bibinfo {author} {\bibfnamefont {J.-Y.}\ \bibnamefont {Courtois}},\
  }\bibfield  {title} {\enquote {\bibinfo {title} {Semiclassical monte carlo
  approaches for realistic atoms in optical lattices},}\ }\href@noop {}
  {\bibfield  {journal} {\bibinfo  {journal} {Eur. Phys. J. D}\ }\textbf
  {\bibinfo {volume} {6}},\ \bibinfo {pages} {29} (\bibinfo {year}
  {1999})}\BibitemShut {NoStop}%
\bibitem [{\citenamefont {Renzoni}(2009)}]{ren09}%
  \BibitemOpen
  \bibfield  {author} {\bibinfo {author} {\bibfnamefont {F.}~\bibnamefont
  {Renzoni}},\ }\href@noop {} {\bibfield  {journal} {\bibinfo  {journal} {Adv.
  At. Mol. Opt. Phys.}\ }\textbf {\bibinfo {volume} {57}},\ \bibinfo {pages}
  {1} (\bibinfo {year} {2009})}\BibitemShut {NoStop}%
\bibitem [{\citenamefont {Schiavoni}\ \emph {et~al.}(2002)\citenamefont
  {Schiavoni}, \citenamefont {Carminati}, \citenamefont {Sanchez-Palencia},
  \citenamefont {Renzoni},\ and\ \citenamefont {Grynberg}}]{renzoni02b}%
  \BibitemOpen
  \bibfield  {author} {\bibinfo {author} {\bibfnamefont {M.}~\bibnamefont
  {Schiavoni}}, \bibinfo {author} {\bibfnamefont {F.-R.}\ \bibnamefont
  {Carminati}}, \bibinfo {author} {\bibfnamefont {L.}~\bibnamefont
  {Sanchez-Palencia}}, \bibinfo {author} {\bibfnamefont {F.}~\bibnamefont
  {Renzoni}}, \ and\ \bibinfo {author} {\bibfnamefont {G.}~\bibnamefont
  {Grynberg}},\ }\bibfield  {title} {\enquote {\bibinfo {title} {Stochastic
  resonance in periodic potentials: Realization in a dissipative optical
  lattice},}\ }\href@noop {} {\bibfield  {journal} {\bibinfo  {journal}
  {Europhys. Lett.}\ }\textbf {\bibinfo {volume} {59}},\ \bibinfo {pages} {493}
  (\bibinfo {year} {2002})}\BibitemShut {NoStop}%
\bibitem [{\citenamefont {Carminati}\ \emph {et~al.}(2003)\citenamefont
  {Carminati}, \citenamefont {Schiavoni}, \citenamefont {Todorov},
  \citenamefont {Renzoni},\ and\ \citenamefont {Grynberg}}]{renzoni03}%
  \BibitemOpen
  \bibfield  {author} {\bibinfo {author} {\bibfnamefont {F.-R.}\ \bibnamefont
  {Carminati}}, \bibinfo {author} {\bibfnamefont {M.}~\bibnamefont
  {Schiavoni}}, \bibinfo {author} {\bibfnamefont {Y.}~\bibnamefont {Todorov}},
  \bibinfo {author} {\bibfnamefont {F.}~\bibnamefont {Renzoni}}, \ and\
  \bibinfo {author} {\bibfnamefont {G.}~\bibnamefont {Grynberg}},\ }\bibfield
  {title} {\enquote {\bibinfo {title} {Pump-probe spectroscopy of atoms cooled
  in a 3d lin-perp-lin optical lattice},}\ }\href@noop {} {\bibfield  {journal}
  {\bibinfo  {journal} {Eur. Phys. J. D}\ }\textbf {\bibinfo {volume} {22}},\
  \bibinfo {pages} {311} (\bibinfo {year} {2003})}\BibitemShut {NoStop}%
\bibitem [{\citenamefont {Sanchez-Palencia}\ and\ \citenamefont
  {Grynberg}(2003)}]{grynpra2003}%
  \BibitemOpen
  \bibfield  {author} {\bibinfo {author} {\bibfnamefont {L.}~\bibnamefont
  {Sanchez-Palencia}}\ and\ \bibinfo {author} {\bibfnamefont {G.}~\bibnamefont
  {Grynberg}},\ }\bibfield  {title} {\enquote {\bibinfo {title}
  {Synchronization of hamiltonian motion and dissipative effects in optical
  lattices: Evidence for a stochastic resonance},}\ }\href@noop {} {\bibfield
  {journal} {\bibinfo  {journal} {Phys. Rev. A}\ }\textbf {\bibinfo {volume}
  {68}},\ \bibinfo {pages} {023404} (\bibinfo {year} {2003})}\BibitemShut
  {NoStop}%
\bibitem [{\citenamefont {Cubero}\ and\ \citenamefont
  {Renzoni}(2016{\natexlab{b}})}]{cubren_prl2016}%
  \BibitemOpen
  \bibfield  {author} {\bibinfo {author} {\bibfnamefont {D.}~\bibnamefont
  {Cubero}}\ and\ \bibinfo {author} {\bibfnamefont {F.}~\bibnamefont
  {Renzoni}},\ }\bibfield  {title} {\enquote {\bibinfo {title} {Hidden
  symmetries, instabilities, and current suppression in brownian ratchets},}\
  }\href@noop {} {\bibfield  {journal} {\bibinfo  {journal} {Phys. Rev. Lett.}\
  }\textbf {\bibinfo {volume} {116}},\ \bibinfo {pages} {010602} (\bibinfo
  {year} {2016}{\natexlab{b}})}\BibitemShut {NoStop}%
\bibitem [{\citenamefont {Kloeden}\ and\ \citenamefont
  {Platen}(1992)}]{kloeden}%
  \BibitemOpen
  \bibfield  {author} {\bibinfo {author} {\bibfnamefont {P.E.}\ \bibnamefont
  {Kloeden}}\ and\ \bibinfo {author} {\bibfnamefont {E.}~\bibnamefont
  {Platen}},\ }\href@noop {} {\emph {\bibinfo {title} {Numerical Solution of
  Stochastic Differential Equations}}}\ (\bibinfo  {publisher} {Springer},\
  \bibinfo {address} {New York},\ \bibinfo {year} {1992})\BibitemShut {NoStop}%
\end{thebibliography}
%


\end{document}